%% file: main-PRL.tex
\documentclass[
  aps,prl,twocolumn,
  preprintnumbers,
  amsmath,amssymb
]{revtex4-1}

\usepackage{graphicx}
\usepackage{dcolumn}
\usepackage{bm}

\usepackage{tikz} 
\usetikzlibrary{calc,arrows.meta}

\usepackage{amsmath,amssymb}
\usepackage{hyperref}
\usepackage{utfsym}
\usepackage{physics}
\usepackage{subcaption}
\usepackage{soul}
\usepackage{package-notes}

\usepackage{xcolor}

\usepackage{comment}

\newcommand{\im}{\mathrm{Im}}

\begin{document}
\preprint{LAPTH--XX/25}
\title{Tracking S-matrix bounds across dimensions
}

\author{Mehmet Asım Gümüş}
\email{mehmet.gumus@lapth.cnrs.fr}
\author{Simon Metayer}%
\email{simon.metayer@lapth.cnrs.fr}
\author{Piotr Tourkine}%
\email{piotr.tourkine@lapth.cnrs.fr}
\affiliation{%
\small{LAPTh, CNRS et Université Savoie Mont-Blanc, 9 Chemin de Bellevue, F-74941 Annecy, France}
}%

\date{\today}

\begin{abstract}
\noindent We study massive $2\!\!\to\!\!2$ scattering of identical scalar particles in spacetime dimensions 
3 to 11 using non-perturbative S-matrix bootstrap techniques. 
Treating $d$ as a continuous parameter, we compute two-sided numerical bounds on low-energy observables and find smooth branches of extremal amplitudes separated by sharp kinks at $d=5$ and $d=7$, coinciding with a transition in threshold analyticity and the loss of some well-known 
dispersive positivity constraints. 
Our results reveal a rich structure in the space of massive S-matrices across dimensions and identify threshold singularities as a key organizing principle. We comment on numerical limitations at large dimension and on possible implications for ultraviolet completion in higher-dimensional quantum field theory.
\end{abstract}

\maketitle

\setcounter{secnumdepth}{3}
\section{Introduction}
\label{sec:intro}

Relativistic scattering amplitudes provide a sharp and universal probe of quantum field theory (QFT). For theories with a mass gap, the $2\to2$ S-matrix is a well-defined observable, constrained by analyticity, crossing symmetry and unitarity (ACU). Being non-perturbative, it is sensitive to both low-energy and the structure of the ultra-violet (UV) completion, and is defined without reference to a Lagrangian.

In $d=4$ spacetime dimensions, there is a clear physical paradigm for a massive relativistic S-matrix. The prototypical example is glueball scattering in confining Yang–Mills theory, where a mass gap and local scattering amplitudes arise from a microscopic, asymptotically free QFT. Beyond four dimensions, this picture becomes increasingly more complex: known microscopic constructions involve a variety of mechanisms, including massless degrees of freedom, strong coupling, gravitational and string-like dynamics. 
Many open questions arise, the prime of which being: is 4d-like, local QFT scattering possible at all in higher dimensions?

Motivated by these questions, in this paper, we begin to chart the space of S-matrices for the $2\to2$ scattering of massive scalars in higher dimensions $3\leq d\leq 11$, including half-integer values.
%
%
The gapped setup provides indeed an especially clean testing ground for this program. Thanks to ACU, low and high energy behavior of the amplitude of massive particles is under good control at various scales. We have the rigorous Froissart bound restricting the growth in the UV in all dimensions~\cite{Maharana:2016epi,Zhiboedov:2021SMatrixNotes,Correia:2025uvc}, while elastic unitarity condition forbids any particle production near the two-particle threshold and allows for diverging scattering lengths. The assumption of lightest particle maximal analyticity allows very strongly constrain the scattering amplitude. Together, these features place strong constraints on the analytic structure of massive amplitudes across a wide range of scales, materializing the motto, "not everything is allowed".

In this paper, we employ a non-perturbative constructive (``primal'') S-matrix bootstrap approach based on semidefinite optimization, which allows ACU to be imposed directly at the level of the amplitude while treating the spacetime dimension as a continuous parameter.

Our main results are two-sided numerical bounds on low-energy observables, defined as derivatives of the non-perturbative amplitude at the crossing-symmetric point. As the spacetime dimension is varied, these bounds evolve smoothly over extended ranges of $d$, but exhibit sharp kinks at $d=5$ and $d=7$, while numerical convergence becomes increasingly challenging for $d\gtrsim 10$. We further correlate these kinks with the onset of specific infrared features near the two-particle threshold, signaling a qualitative change in how ACU is realized for massive S-matrices beyond certain critical dimensions.

The  paper is organized as follows. In Section~\ref{sec:setup}, we define our bootstrap setup and the low-energy observables of interest. Section~\ref{sec:threshold} discusses the role of two-particle threshold behaviour and its implications for dispersive representations. Our numerical results are presented in Section~\ref{sec:results}, and their interpretation is discussed in Section~\ref{sec:discussion}. Finally, an outlook is given in Section~\ref{sec:outlook}.

\section{Setup}
\label{sec:setup}

We consider Lorentz-invariant $2\to2$ scattering amplitudes $A(s,t,u)$ for identical massive scalar particles in arbitrary spacetime dimensions $d$ without cubic interactions. The amplitude is a function of the Mandelstam invariants
\begin{equation}
s=(p_1+p_2)^2,\quad t=(p_1-p_3)^2,\quad u=(p_1-p_4)^2,\,
\end{equation}
with $s+t+u=4$, where we work in units where the particle mass is set to $p_i^2=m^2=1$.

As is standard, the amplitude is assumed to satisfy analyticity, crossing symmetry, and unitarity. We impose \textit{maximal} analyticity, namely analyticity in the complex variables $s,t,u$ away from the physical branch cuts $s,t,u\ge4$ and crossed-cuts. Crossing symmetry for identical scalars implies full symmetry under permutations of $(s,t,u)$.

For physical $s$-channel kinematics ($s\ge4$, $t\le0$), the amplitude admits the partial-wave expansion
\begin{equation}
A(s,t)=\sum_{\substack{J=0\\ J~\mathrm{even}}}^{\infty} n_J^{(d)}\, f_J(s)\, P_J^{(d)}(z),
\quad 
z=1+\frac{2t}{s-4},
\label{eq:PWE}
\end{equation}
where $P_J^{(d)}(z)$ are appropriately normalized Gegenbauer polynomials and $n_J^{(d)}$ are the corresponding normalization factors and odd spins $J$ drop for identical external scalars. We follow the conventions of~\cite{Correia:2020xtr} and explicit expressions and conventions are collected in Appendix~\ref{app:conventions}.

Unitarity is imposed at the level of partial waves $f_J(s)$. 
Defining
\begin{equation}
S_J(s)=1+i\,\phi_d^2(s)\,f_J(s),
\end{equation}
with $\phi_d^2(s){=}\sqrt{s-4}^{d-3}\!/\sqrt{s}$ the $d$-dimensional two-particle phase space factor, unitarity implies
\begin{equation}
|S_J(s)|^2\le1 ,
\label{eq:unitarity}
\end{equation}
for all $s\ge4$ and even spins $J$. Between the 2- and 4-particle thresholds, $S_J$ satisfies the stronger condition of \textit{elastic} unitarity
\begin{equation}
|S_J(s)|^2= 1 ,\quad 4\leq s\leq 16\,.
\label{eq:elastic-unitarity}
\end{equation}

In this letter, we study standard, non-perturbative low-energy observables defined by values of the scattering amplitude around the cross-symmetric point $s=t=u=4/3$:
\begin{equation}
\bar c_0=\frac{\mathcal{N}_d}{2}\,A\!\left(\tfrac{4}{3},\tfrac{4}{3}\right),
\quad
\bar c_2=\frac{\mathcal{N}_d}{8}\,
\left.\frac{\partial^2}{\partial s^2}A(s,\tfrac{4}{3})\right|_{s=4/3}.
\label{eq:c0-c2-def}
\end{equation}
The normalization with $\mathcal{N}_d \sim \Gamma(d/2)^{-1}$ can be motivated on general grounds by the growth of the amplitude, so that the $\bar c$'s are of order 1 at large $d$. We describe this together with a generic bound on the amplitude at large $d$ from a straightforward saddle point analysis in Appendix~\ref{app:larged}.
At low energies, the amplitude admits an effective-field-theory expansion around this point,
\begin{equation}
A(s,t){=}
c_0 + c_2\!\left(\bar s^2{+}\bar t^2{+}\bar u^2\right) + c_3 (\bar{s}\bar{t}\bar{u})
+c_4\!\left(\bar s^2{+}\bar t^2{+}\bar u^2\right)^{\!2}
\!+\dots,
\label{eq:wilsons}
\end{equation}
for $(\bar s,\bar t,\bar u)=(s-4/3,t-4/3,u-4/3)$ small and where $c_i$'s are the Wilson coefficients, related to the properly normalized ones by $\bar c_i = \mathcal{N}_d c_i/2$.

Our numerical strategy follows the approach of~\cite{Paulos:2016fap,Paulos:2016but,Paulos:2017fhb}, further developed in~\cite{Guerrieri:2021ivu,EliasMiro:2022xaa}.
Numerical bounds on $\bar c_0$ and $\bar c_2$  coefficients are obtained by optimizing them over the space of amplitudes satisfying ACU. 
We start with a crossing-symmetric, maximally analytic ansatz for the amplitude. Unitarity \eqref{eq:unitarity} is then imposed as a semi-definite positivity constraint, with a truncation in spin and over a large grid of values of $s$, which converges to the bound as the number of parameters in the ansatz, spin truncation and grid size become large.
The details of the ansatz, the numerical grids, and the extrapolation procedure are described in Appendix~\ref{app:primal_bootstrap}. 

\section{Threshold behavior and dispersive constraints}
\label{sec:threshold}

In the defining equations of the last section, the dimension $d$ appeared analytically. However, one important outcome of our analysis is that one cannot be completely agnostic about it. In the UV, the Froissart bound holds in all $d$, ensuring that two subtractions at infinity are enough, irrespective of $d$. However, in the infra-red it turns out that the threshold behavior can generate divergences at the two-particle threshold in dispersion relations, causing a loss of positivity of certain coefficients.

Let us explain briefly the main idea.
As described in ref.~\cite[sec.~5]{Correia:2020xtr}, elastic unitarity \eqref{eq:elastic-unitarity} constrains the behavior of {partial waves} near $s=4$. It allows for singular threshold behavior up to the most divergent power 
$    (s-4)^{-(d-3)/2}\,,$
with logarithmic correction for odd integer $d$. In the following, we refer to threshold singularities $(s-4)^{-n}$ as \textit{mild} or  \textit{strong}, depending on $n$:
\begin{equation}
    \frac1{(s-4)^n}\Leftrightarrow 
    \begin{cases}
       \,\, n\leq1 \rightarrow \text{``mild" singularity}\\
       \,\, n>1 \rightarrow \text{``strong" singularity}
    \end{cases}\,,
\end{equation}
with logarithmic corrections in odd dimensions, such that for instance $\frac1{(s-4) \log(4-s)}$ is a \textit{mild} singularity in $d=5$.
Note that we do include $n=1$ as a mild singularity: the reason will become clear later.

These divergences, while allowed by elastic unitarity, have a direct impact on dispersion relations for $d\geq5$. The singular threshold terms render the amplitude non-integrable near the two-particle threshold inside dispersion integrals, spoiling standard positivity arguments and allowing for \emph{threshold-induced negativity}. To our knowledge, this effect has not been previously discussed.
Depending on $d$, some low-energy coefficients then become genuine \textit{IR subtraction constants}. As a result, the observables $c_{2n}\sim \partial_s^{2n}A(s,4/3)|_{s=4/3}$ are no longer sign-definite for $2n\leq d-4$. Positive combinations nevertheless still exist; for instance,
\begin{equation}
\begin{aligned}
   d=4,&\quad c_2  \geq 0,\\
   d=6,&\quad c_2 - \tfrac{128}{9} c_4 \geq0, ~ \dots
    \end{aligned}
\end{equation}

In conclusion, including the threshold terms in the ansatz is essential to capture the full allowed space, primarily because the negative regions cannot be reached otherwise.
But the threshold also affects the numerical convergence, even when negativity is not present. The speed-up obtained by an ansatz augmented by the singular pieces in $d=4$ (where the threshold singularity is integrable) was noted long ago in~\cite{Paulos:2017fhb}. In higher-$d$, the situation is worse: not just the leading but the full set of subleading singularities needs to be included with care to obtain convergence. The complete construction of our threshold ansatz, including subtleties specific to odd dimensions, is detailed in Appendix~\ref{app:art-TH}, and the IR subtractions in Appendix~\ref{app:ir_sub}.

\section{Results}
\label{sec:results}

In this section we present numerical bounds obtained from the primal bootstrap described in Section~\ref{sec:setup} across dimensions $3\leq d\leq 11$, including half-integer values. We first present bounds on the rescaled low-energy observables $\bar c_0$ and $\bar c_2$. We then focus on the detailed structure of the space of admissible amplitudes in $d=6$, both without and with an additional bound-state pole, as a test of the threshold structure. 
The main qualitative outcome of the results shown below is the emergence of two smooth families of extremal amplitudes, separated by sharp kinks whose origin we trace to threshold structure, detailed in the Table~\ref{tab:th-horizontal}.

\paragraph{Numerics.} 
A few comments on numerics are in order before we start. Heavy automatization was required to produce results across that many dimensions. The main bottlenecks were the evaluation of a large number of integrals (partial-wave projections), large-scale data management, and the convex optimization of very large systems of inequalities. Integrals are computed with \textit{Mathematica} and stored in an SQL database for efficient reuse, while the resulting semi-definite programs are solved using SDPB~\cite{Simmons-Duffin:2015qma,Landry:2019qug}. 
All stages of the computation are carried out using grid computing resources on several super-computing facilities. A \textit{Mathematica} wrapper was developed to handle these tasks efficiently~\cite{Metayer:SDPBwrapper}.
In total, this project required computing ${\sim}10^7$ integrals and ${\sim}5000$ SDPB points, for a total of ${\sim}80,000$ CPU hours, producing a database of ${\sim}100$Gb. 

As for numerical convergence, our numerical procedure involve cutoffs in energy, spin and ansatz size. However, convergence in spin is essentially exact due to the use of subtracted positivity relations imposed as extra linear constraints (see Appendix~\ref{app:primal_opt_prob}), and variations of the energy grid sampling and of the ultraviolet cutoff do not lead to qualitative changes in the results. Therefore, the only numerical cutoff left is ansatz size $N_{\max}$, where $\#\text{params} \sim N_{\max}^2$. In the following, we stop at $N_{\max}=20$ for all dimensions, except for $d=11$, where we have to go to $N_{\max}=29$ to obtain reasonable bounds.

All results presented below include the complete threshold structure allowed by elastic unitarity, as described in Section~\ref{sec:threshold} and listed in Appendix~\ref{app:threshold}, and the optimizer decides for which solutions to turn singular terms on or off.

\begin{figure*}[t]
    \centering
    \begin{subfigure}{0.45\textwidth}
        \centering
        \includegraphics[width=0.95\linewidth]{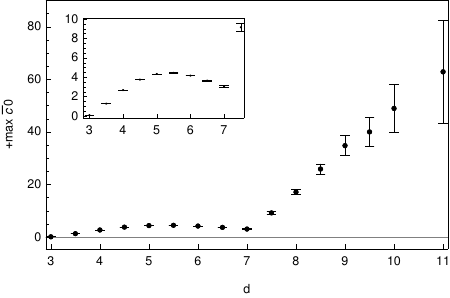}
        \vspace{-0.25cm}
        \caption{$+\max \bar c_{0}(d)$}
        \label{fig:c0max}
    \end{subfigure}
    \begin{subfigure}{0.45\textwidth}
        \centering
        \includegraphics[width=0.95\linewidth]{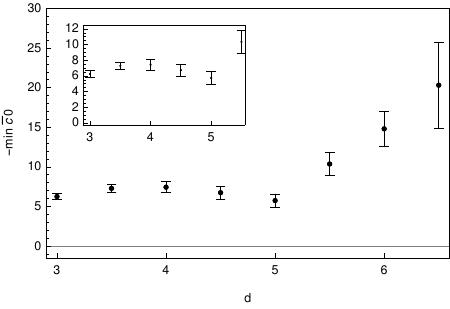}
        \vspace{-0.25cm}
        \caption{$-\min \bar c_{0}(d)$}
        \label{fig:c0min}
    \end{subfigure}
    \vspace{1em}
    \begin{subfigure}{0.45\textwidth}
        \centering
        \includegraphics[width=0.95\linewidth]{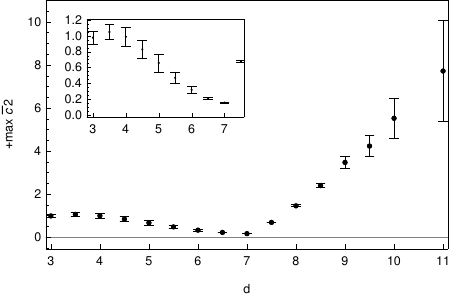}
        \vspace{-0.25cm}
        \caption{$+\max \bar c_{2}(d)$}
        \label{fig:c2max}
    \end{subfigure}
    \begin{subfigure}{0.45\textwidth}
        \centering
        \includegraphics[width=0.95\linewidth]{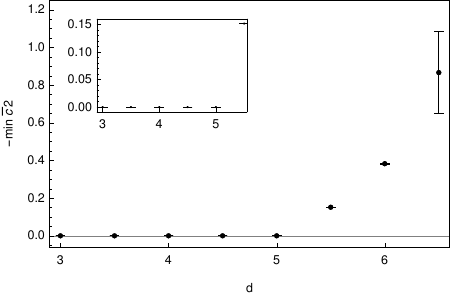}
        \vspace{-0.25cm}
        \caption{$-\min \bar c_{2}(d)$}
        \label{fig:c2min}
    \end{subfigure}
    \vspace{-0.5cm}
    \caption{Extrema of $\bar c_0$ and $\bar c_2$ as functions of $d$. The sharp kinks visible at $d=5$ and $d=7$ coincide with a qualitative change in the threshold structure of the corresponding extremal amplitudes (see Discussion~\ref{sec:discussion}).
    Lower error bars corresponds to the best values obtained at finite truncation ($N_{\max}=20$, except in $d=11$ where $N_{\max}=29$), while the upper error bars shows the extrapolated estimates obtained from a linear fit in $(1/N_{\max},\bar c_i)$ using the last ten data points. Variations of the fitting procedure lead to qualitatively similar results (see Appendix~\ref{app:convergence} for more).}.
    \vspace{-0.5cm}
    \label{fig:fourpanel}
\end{figure*}

\vspace{-0.25cm}
\subsection{Bounds on $\bar c_0$ and $\bar c_2$}
\vspace{-0.25cm}

We determine numerical upper and lower bounds on the low-energy coefficients $\bar c_0$ and $\bar c_2$, defined in eq.~\eqref{eq:c0-c2-def}, as functions of the spacetime dimension $d$. The resulting bounds for $3\le d\le11$ are shown in Figs.~\ref{fig:c0max}--\ref{fig:c2min}. For both coefficients, the extremal values vary continuously with $d$ over extended ranges, with half-integer dimensions smoothly interpolating between neighboring integers.

Strikingly, the extremal curves exhibit sharp kinks at specific dimensions. For both $\bar c_0(d)$ and $\bar c_2(d)$, the maximal bound displays a kink at $d=7$, while the minimal bound exhibits a corresponding feature at $d=5$, which coincides with $\bar c_2$ becoming negative. These features persist as the truncation parameters are increased and are stable under variations of the numerical setup. For $d\le10$, convergence is reliable as the spin and energy truncations are increased, becoming slower but remaining under control above the kinks; for $d=11$, we require $N_{\rm max}=29$ to obtain a more realistic picture.

Remarkably, both coefficients follow the same generic pattern, therefore the appearance of the kinks reflects a reorganization of the space of ACU-consistent S-matrices significant enough to affect simultaneously multiple, a priori uncorrelated low-energy observables.

\paragraph{Structure of extremal solutions across kinks.}

The appearance of kinks in the bounds is accompanied by qualitative changes in the threshold structure of the corresponding extremal solutions, as summarized in Table~\ref{tab:th-horizontal}.
\begin{table}[h!]
\centering
\small
\setlength{\tabcolsep}{4pt}
\renewcommand{\arraystretch}{1.2}
\begin{tabular}{c|cccccccc}
\hline
$d$ 
& 3 & 4 & 5 & 5.5 & 6 & 6.5 & 7 & 7.5--10 \\
\hline
\multicolumn{9}{c}{$\bar c_0^{\,\max}$ and $\bar c_2^{\,\max}$ threshold terms} \\
\hline
$n \leq 1$ & ✗ & ✓ & ✓ & ✓ & ✓ & ✓ & ✓ & ✓ \\
$n>1$ & ✗ & ✗ & ✗ & ✗ & ✗ & ✗ & ✗ & ✓ \\
\hline
\multicolumn{9}{c}{$\bar c_0^{\,\min}$ and $\bar c_2^{\,\min}$  threshold terms} \\
\hline
$n\leq1$ & ✗ & ✓ & ✓ & ✓ & ✓ & ✓ & ✓ & ✓ \\
$n>1$ & ✗ & ✗ & ✗ & ✓ & ✓ & ✓ & ✓ & ✓ \\
\hline
\end{tabular}
\caption{
Activation of threshold singularities in extremal amplitudes, indicating for each $d$ the activation of $(s-4)^{-n}$ terms with $n\leq1$ or $n>1$ (or $(s-4)^{-n}\log(4-s)^{-p}$ in odd dimensions). Check-marks denote nonzero coefficients within numerical tolerance.
}
\label{tab:th-horizontal}
\end{table}

\vspace{-0.7cm}
\subsection{Space of amplitudes and cubic coupling in $d=6$}
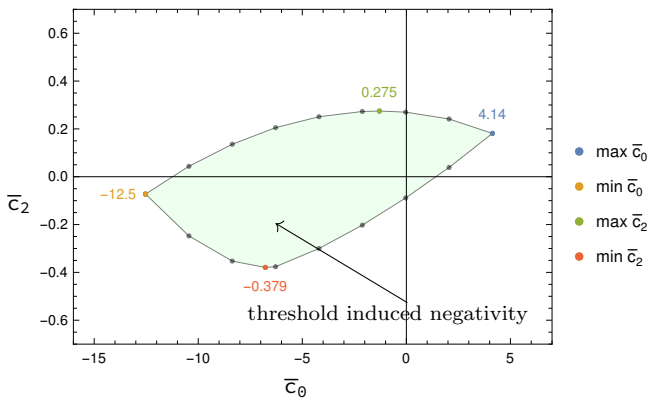
\begin{figure}[h!]
    \centering
    \begin{tikzpicture}
        \node[inner sep=0] (img) {\includegraphics[width=1\linewidth]{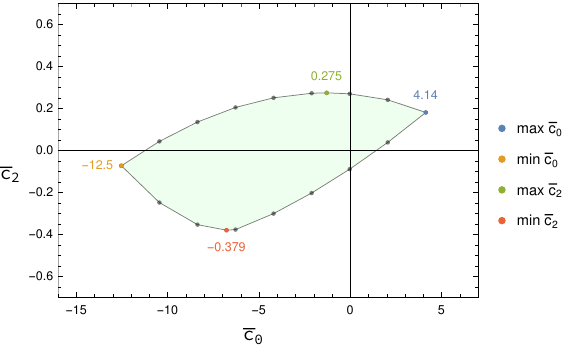}};
                \node at ($ (img.south west)!0.75!(img.south east) !0.22! (img.north west) $) {\footnotesize threshold induced negativity};
        \coordinate (A) at ($ (img.south west)!0.82!(img.south east) !0.25! (img.north west) $);
        \coordinate (B) at ($ (img.south west)!0.75!(img.south east) !0.45! (img.north west) $);
        \draw[->] (A) -- (B);
    \end{tikzpicture}
    \caption{Allowed region in the $(\bar c_0,\bar c_2)$ plane in $d=6$.
    }
    \vspace{-0.5cm}
    \label{fig:almond_d_6}
\end{figure}

\vspace{-0.1cm}
In order to describe more precisely the structure of leading and  subleading threshold terms, we now focus on a fixed dimension: $d=6$. The full allowed region in the $(\bar c_0,\bar c_2)$ plane is shown in Fig.~\ref{fig:almond_d_6}. This region is convex and bounded, with extremal points corresponding to the solutions saturating the bounds discussed above. It is similar to regions obtained in $d\leq4$~\cite{Chen:2022nym,Gumus:2023xbs}, except for the notable threshold-induced-negativity of $\bar c_2$.

\paragraph{Bound-state pole and cubic coupling.}

In $d=6$, the cubic coupling $\phi^3$ is marginal. For this reason, we momentarily extend the scope of the paper and relax the no-cubic-coupling assumption, in order to allow for the presence of a bound-state pole,
\begin{equation}
A(s,t) \supset \frac{g^2}{s-m_b^2} + \text{crossed}\,,
\end{equation}
with $1 \le m_b^2 \le 4$. The case $m_b^2=1$ corresponds to a self-coupling of the external particle.

\begin{figure}[h!]
    \centering
    \includegraphics[width=0.9\linewidth]{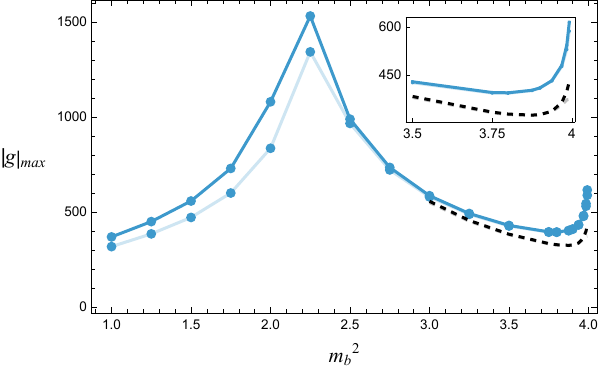}
    \definecolor{darkblue}{RGB}{42, 112, 153}
    \definecolor{lightblue}{RGB}{145, 178, 197}
    \vspace{-0.25cm}
    \caption{Bounds on the residue of a bound-state pole in $d=6$ as a function of the bound-state mass $m_b^2$. The two shades of blue indicate $N_{\max}=8$ ({\color{lightblue}\textbf{lighter}}) and $N_{\max}=10$ ({\color{darkblue}\textbf{darker}}). Dashed black (\textbf{- - -}) shows the renormalised coupling $g_{\text{ren}}$ described in the text, which converges to a finite value at $m_b^2=4$.}
    \label{fig:pole_d_6}
\end{figure}

We determine upper bounds on the magnitude of the cubic coupling $|g|$ as a function of the bound-state mass $m_b^2$, as the similar analysis of \cite{Paulos:2017fhb,Guerrieri:2023qbg} in $d=4$. The resulting bounds are shown in Fig.~\ref{fig:pole_d_6}. The maximal coupling exhibits a peak around $m_b^2\simeq2$, followed by a divergence as $m_b^2\to4$.
The divergence near threshold is easy to understand: since the ansatz possesses a  threshold singular term $\frac{\beta}{s-4}$ with free coefficient $\beta$ (see eq.\eqref{eq:th-subth-all}), as $m_b^2$ approaches $4$, it is easy to make $|g|$ arbitrarily big by tuning $\beta$ accordingly. Hence, we define a renormalized coupling $g_{\rm ren}^2 = g^2 -\beta $ as $s\to4^-$, which we observe to converge to a finite value. We show this $g_{\rm ren}$ in Fig.~\ref{fig:pole_d_6}.

\section{Discussion}
\label{sec:discussion}

Having presented the numerical results, let us now discuss their implications. For both the minimal and maximal extremal amplitudes, we observed the existence of two distinct branches of solutions, separated by a sharp kink at $d=5$ and $d=7$, respectively. Along each branch, the extremal amplitudes vary smoothly with the spacetime dimension.
These smooth branches are numerically robust and persist over extended ranges of $d$. Their smoothness suggests that, within each branch, a common mechanism produces ACU-consistent S-matrices, while the kink signals a reorganization of the structure of the amplitudes. Understanding the origin of this reorganization is therefore central to interpreting the bounds.

\paragraph{Kink: mild versus strong threshold behavior.}
While we cannot predict \emph{a priori} the dimension at which a kink should occur, we find empirically that the kink coincides with the onset of strong threshold behavior in the extremal solutions. Concretely, amplitudes before and after the kink differ by the activation of threshold singularities of the form $(s-4)^{-n}$ with $n>1$. Evidence for this correspondence is summarized in Table~\ref{tab:th-horizontal}. 

The most convincing signal appears on the maximal branch. Although strong threshold divergences are allowed by elastic unitarity for $d>5$, the extremal solutions along the smooth branch systematically avoid them: only mild threshold singularities are selected, even when stronger ones are kinematically allowed. To be fully concrete, in $d=6$ for instance, singularities $(s-4)^{-3/2},(s-4)^{-1},(s-4)^{-1/2}$ are allowed (see \eqref{eq:th-subth-all}): on the maximal solutions, the first one is turned off and the latter ones are turned on.
At the kink, strong threshold contributions are activated and remain present beyond it. This motivates the terminology
\begin{center}
\textit{mild threshold branch}\quad versus \quad
\textit{strong threshold branch} 
\end{center}
for the branches before and after the kink, respectively, on either maximal/minimal $\bar c_i$ solutions.
The coincidence of the kinks in $\bar c_0$ and $\bar c_2$ with this transition is nontrivial. In particular, the extremization of $\bar c_0$ does not generically coincide with that of $\bar c_2$, see e.g., Fig.~\ref{fig:almond_d_6} in $d=6$.

On the minimal $\bar c_0$ or $\bar c_2$ solutions, the situation is more transparent. At $d=5$, the dispersive representation of $\bar c_2$ breaks down due to threshold effects, and $\bar c_2$ ceases to be positive-definite. This directly explains the kink observed in $\bar c_2^{\rm min}$ in Fig.~\ref{fig:c2min}. However, since $\bar c_0$ was never dispersive or sign-definite, the fact that the kink also appears in $\bar c_0^{\rm min}$ at $d=5$ remains unexplained.

We do not have an analogous understanding of the origin of the $d=7$ kink for the max solutions. Importantly, these solutions illustrate that the distinction between mild and strong threshold behavior is \textit{not} tied to integrability at threshold: non-integrable terms such as $(s-4)^{-1}$ already appear in $d\leq5$ for these maximal amplitudes. The $d=7$ kink therefore reflects a more specific reorganization of the threshold expansion.

Overall, the simultaneous appearance of kinks across different observables, correlated with the onset of strong threshold behavior, indicates that a genuine structural effect is at work. We leave a more detailed analysis of the resulting amplitudes for future companion study~\cite{Gumus:2026WIP}.

\paragraph{Subleading threshold terms and bound-states in $d{=}6$.}
We further studied cubic couplings to a bound state below threshold in $d=6$. This makes physical sense as $\phi^3$ is marginal in $d=6$, and constitutes a relevant data-point for further higher-$d$ studies. In addition, it sheds light on a new feature compared to $d=4$, where in particular it was observed~\cite{Paulos:2017fhb,Guerrieri:2023qbg} that such couplings decouple at the threshold $m_b^2\to4$. This is consistent with the fact that the worst singularity allowed at $s=4$ by elastic unitarity is a milder singularity, $(s-4)^{-1/2}$. In $d=6$, a pole $(s-4)^{-1}$ is allowed: consistently, we found that the coupling of this term goes to a finite value when $m_b^2\to4$.

\paragraph{Numerical complexity : min versus max;  large-$d$.}
For minimized observables, convergence is much more difficult than for maximal. This is a known effect in $d=4$~\cite{Paulos:2017fhb,Guerrieri:2021tak,EliasMiro:2022xaa,Correia:2025uvc}, see also the older Ref.~\cite{Auberson:1979ye}. It has been hypothesized in \cite{Guerrieri:2023qbg} that this is related to a spin 2 resonance trying to become a bound-state, which is difficult to achieve numerically because this implies that amplitude needs to Reggeize in order not to violate Froissart growth. We see this effect in all $d$, to the point that our numerics actually becomes unable to capture accurate bounds past $d=6$ and will require a different approach to probe this region. Introducing a spin $>2$ threshold divergence following \cite{Auberson:1979ye} has the potential to help the numerics.

For the max amplitudes, we have better convergence, and can reach $d=10$, and even $d=11$ but at the high cost of doubling the amount of terms in the ansatz. Generally, extrapolations become less precise and can only catch a rough trend at higher $d$.
While convergence in spin remains under control, one obstruction appears to be of infrared in nature. The allowed threshold  structures become increasingly singular with growing $d$, and even when these contributions are explicitly included, the present wavelet-based ansatz struggles to efficiently capture the resulting IR complexity. This points to the need for more flexible ansatz, in particular with a refined treatment of threshold behavior and general analytic structure.
 
\vspace{-0.5cm}
\section{Outlook}
\label{sec:outlook}
\vspace{-0.25cm}

In this paper, we presented the results of an exhaustive bootstrap analysis for gapped S-matrices in dimensions $d>4$. To our knowledge, this is the first systematic exploration of the kind.
Let us conclude with a few speculative remarks and open questions.

\paragraph{Two families of massive S-matrices.}
Beyond the kinks identified above, the extremal amplitudes display strong threshold behavior, including higher-order singularities such as double poles. While such structures are not excluded by general locality or analyticity principles, they cannot arise from the exchange of a single local elementary degree of freedom. 
That being said, most of the threshold structures do not come simple perturbative terms, unlike the pole $(s-4)^{-1}$. For instance, in $d=3$, resummation effects generate the threshold divergence in $\phi^4$~\cite{Bros:1998tt}. On general grounds, threshold divergences are thus related to resummation effects.
From that perspective, the distinction between strong and mild nevertheless indicate two qualitatively distinct physical phenomena are at stake in the distinction: $n\leq 1$ and $n>1$. Even within the resummation discussion, because simple poles are the fastest growing singularities of the ``mild" class, it is tempting to suggest that the kink might have a link with the onset of some form of \textit{non-locality}, at minimum reflecting stronger non-elementary dynamics in higher dimensions.
Furthermore, it is worth to note that the existence of a smooth branch passing through $d=4$ shows that gapped ACU-consistent S-matrices in four dimensions belong to a continuous family extending to higher dimensions. This raises the possibility that $4d$-like scattering persist beyond four dimensions, for instance as genuinely higher-dimensional asymptotically safe theories, or as effective descriptions—possibly string-theoretic in nature—valid over a parametrically large energy range, as in~\cite{Chang:2014jta}.

\paragraph{Relation to stringy and gravitational expectations.}
From a string-theoretic perspective, it is indeed in general expected that UV completion beyond $d=4$ require non-local ingredients related to Little String theories, and that beyond $d=6$ gravity cannot even be fully decoupled~\cite{Maldacena:1997re,Itzhaki:1998dd}, with a possible borderline case in $d=7$ \cite{Itzhaki:1998dd}~\footnote{We thank Miguel Montero for bringing this point to our attention.}. While our analysis does not rely on any gravitational input, it is surprising that the onset of strong threshold behavior for the maximum branch occurs in dimensions where such expectations become relevant. 
The bootstrap we use cannot dynamically generate new light degrees of freedom such as poles or massless cuts, that have to be included by hand. Therefore, it is a possibility that would be important to investigate that the onset of "non-locality" in the form of strong threshold could be a signal that the algorithm is missing some IR structure, such as massless particle exchange.
We also note that gravitational interactions are known to generate negativity Wilson coefficients\cite{Caron-Huot:2021rmr,Caron-Huot:2022ugt,Chang:2025cxc}.
Whether these coincidences reflect a genuine connection between threshold singularities and the necessity of non-local UV structure, or is merely accidental, is a very interesting question, that will require further investigations.

\paragraph{Positivity program with threshold-induced-negativity.} The presence of two-particle threshold divergences for $d>5$ might also impact the positivity program~\cite{Caron-Huot:2020cmc,Bellazzini:2020cot,Arkani-Hamed:2020blm,Tolley:2020gtv,deRham:2022hpx} in situations where IR branch cuts are non-negligible, such as having multiple species~\cite{deRham:2025htd}. Due to this divergence, an infinite number of null constraints would need modification, making the positivity constraints possibly less restrictive in higher dimensions. It would be interesting to study this possibility.


\paragraph{Large-$d$ limit.}

To motivate the scaling of the amplitude, we have observed that, at fixed spin $J$, the partial-wave amplitudes admit a simple saddle-point structure at large spacetime dimension, as discussed in Appendix~\ref{app:larged}. 
At the same time, extending this observation to the full scattering amplitude $A(s,t)$ is highly nontrivial. The large-$d$ limit and partial-wave sum do not commute, preventing a straightforward reconstruction of the amplitude from fixed-$J$ data. This obstruction currently limits the usefulness of the saddle-point analysis to the statement of the boundedness of the amplitude for physical kinematics.
Within the range accessible to our numerics, the strong-threshold branch of maximal amplitudes exhibits a growth with $d$ that appears in tension with the $\mathcal{O}(1)$ scaling expected from these fixed-$J$ considerations.
It is therefore an open question whether ACU-consistent massive S-matrices exist at arbitrarily large spacetime dimension (especially beyond $d=11$), and whether the notion of a large-$d$ bootstrap would be useful or not. Resolving this issue will likely require either new analytic control over the interplay of large $d$ and large spin, and a better numerical control over these large $d$ amplitudes, and maybe open the way to the development of techniques similar those of~\cite{Emparan:2013moa,Emparan:2025yfy} in gravity.

\paragraph{Dual bootstrap.} 
A solid strategy to improve the accuracy of the bounds would be to perform \textit{dual} bootstrap in higher-$d$ as was done in $d=4$ \cite{Guerrieri:2021tak,Gumus:2023xbs}. Dual method cannot access the full amplitude compared to primal, but has the advantage that it produces \textit{rigorous}, strict upper bounds. It would require projecting higher-$d$ dispersion relations on individual partial waves being the primal variables as in \cite{Bhat:2023puy}.

\paragraph{Including inelastic effects.}
Last, but not least, as in other primal bootstrap studies based on the semi-definite approach, our analysis enforces unitarity only in the elastic $2\to2$ channel, and converges towards apparently fully elastic amplitudes $|S_J(s)|=1$ for all $J$ and $s\geq 4$. This is in known tension with general results on particle production at high energies~\cite{Aks:1965qga}. It would be important to investigate how the bounds obtained here are modified once inelastic effects are incorporated, for instance following the approaches developed in~\cite{Tourkine:2021fqh,Tourkine:2023xtu,Gumus:2024lmj}.

\onecolumngrid
\medskip

\begin{center}  
-------------------------------- 
\end{center}

\newpage

\begin{acknowledgments}
\textit{Acknowledgements.} We would like to thank Andrea Guerrieri, Damien Leflot, Slava Rychkov, Xi Yin and Sasha Zhiboedov for useful discussions and comments. This work was granted access to the HPC resources of IDRIS under the allocation 2025-104798 made by GENCI. We thank Marina Cagliari for informing us we could access and helping us setting-up our access to the HPC IDRIS cluster. 
The authors thank the Yukawa Institute for Theoretical Physics at Kyoto University for hosting the meeting
``Progress of Theoretical Bootstrap” where parts of this work were finalized.
This work has received funding from Agence Nationale de la Recherche (ANR), project ANR-22-CE31-0017.
\end{acknowledgments}
\bibliography{bib-papers}
\input{new_appendices.tex}


\end{document}

%% file: new_appendices.tex
\newpage

\section*{Appendices}

\appendix
\setcounter{secnumdepth}{3}

\section{Conventions and normalizations}
\label{app:conventions}

%
In this article, we follow the conventions and normalization choices of \cite{Correia:2020xtr}. 
The Gegenbauer polynomials in $d$ spacetime dimensions are given by
\begin{equation}
\qquad
P^{(d)}_{J}(z)={}_2F_1\left(-J,d+J-3,\frac{d-2}{2},\frac{1-z}{2}\right) \,. 
\label{eq:Pjd-def}
\end{equation}
They constitute a complete orthogonal basis of spherical harmonics in arbitrary dimensions, and are related to the Gegenbauer polynomials $C^{(a)}_n(x)$ as follows
\begin{equation}
    P^{(d)}_J(z)=C^{(d-3)/2}_J(z)/C^{(d-3)/2}_J(1) \, .
\end{equation} 
Inverting the partial wave expansion~\eqref{eq:PWE} using orthogonality of the Gegenbauer polynomials yields the partial wave coefficients
\begin{equation}
f_J(s)=\frac{\mathcal{N}_d}{2}\int_{-1}^{1}\,(1-z^2)^{\frac{d-4}{2}}\,P^{(d)}_{J}(z)\,A(s,t(s,z))\,\mathrm{d}z\,,\qquad t(s,z)=-\frac{1}{2}(s-4)(1-z) \, .
\label{eq:PWProj}
\end{equation}
The two formulas, \eqref{eq:PWE} and \eqref{eq:PWProj}, 
include conventional normalization factors given by
\begin{equation}
n^{(d)}_{J}=\frac{(4\pi)^{d/2}(d+2J-3)\Gamma(d+J-3)}{\pi\,\Gamma\left(\frac{d-2}{2}\right)\Gamma(J+1)}\,,
\qquad
\mathcal{N}_d=\frac{(16\pi)^{\frac{d-2}{2}}}{\Gamma\left(\frac{d-2}{2}\right)}\,.
\label{eq:n-def}
\end{equation}
that stem from the orthogonality relations between the Gegenbauer polynomials. Let us also define 
\begin{equation}
\label{eq:nd}
    n_d = 2^{5-d}n^{(d)}_0 \,,
\end{equation}
for later convenience in the numerical setup section \ref{app:primal_bootstrap}.
%
%
%
Finally, let us recast the unitarity condition \eqref{eq:unitarity} in terms of the partial wave coefficients
\begin{equation}
|S_J(s)|^2\leq 1\quad \Leftrightarrow \quad 2 \,  \im f_J(s)\geq \phi^2_d(s) \, |f_J(s)|^2 \, .
\label{eq:unitarity2}
\end{equation}

\section{Threshold engineering}
\label{app:threshold}

In this section, we review how \emph{elastic} unitarity~\eqref{eq:elastic-unitarity} fixes the threshold expansion of the partial wave coefficients in powers of $s-4$, following~\cite{Correia:2020xtr}.
The $f_J(s)$ are allowed to be singular as $s\to4^+$, and our goals are i) to isolate which singular structures appear in the expansion, independently of any ultraviolet completion or locality assumption ii) to engineer an ansatz term $A_\text{th}(s,t)$ featuring the prescribed singularities with a free coefficient in the primal ansatz \eqref{eq:ansatz}, satisfying the analyticity and crossing-symmetry assumptions.

This classification of singularities plays a central role in the interpretation of the numerical results presented in the main text, in particular the emergence of kinks and the loss of positivity of certain low-energy coefficients in $d \geq 5$ as discussed in~\ref{app:ir_sub}.
%
%
To unclutter the expressions in this section, we also set
\newcommand{\al}{\alpha}
\begin{equation}
    \al \equiv \frac{d-3}2 \,,
\end{equation}

\subsection{Elastic unitarity} 
Elastic unitarity is the statement that, due to the lack of $2 \to $ more processes below the four-particle threshold, the inequality in \eqref{eq:unitarity2} has to become an equality as follows
\begin{equation}
2\im f_J(s) = \frac{(s-4)^{\alpha}}{\sqrt{s}}|f_J(s)|^2\,,\quad 4\leq s\leq 16 \, .
\label{eq:unitarity_el}
\end{equation}
It can be solved for $1/\im f_J(s)$ by using $\im(1/z) = -\im(z)/|z^2|$ and reads
\begin{equation}
\label{eq:el-un-inverse}
2\im \frac{1}{f_J(s)} = -\frac{(s-4)^{\al}}{\sqrt{s}} \,,\quad4\leq s\leq 16\,,
\end{equation}
while $1/\im f_J(s)=0$ for $0 < s < 4$ by real analyticity.
%
Remembering that $2i \Im g(x) = g(x+i\epsilon)-g(x-i\epsilon)$ for a real analytic $g(x)$, we are now searching for a function, whose discontinuity is given by the RHS of \eqref{eq:el-un-inverse}.

For $J\geq2$, the Froissart-Gribov (FG) projection formula (obtained by analytically continuing $f_J(s)$ in spin -- see Eq. (2.53) of \cite{Correia:2020xtr}) implies that partial waves have to vanish near the threshold $f_J(s)\sim (s-4)^J$. We therefore restrict ourselves to $J=0$, for which the FG projection does not hold (since the amplitude is allowed to grow as $s\log(s)^2$).
To find $f_0(s)$, we could write a dispersion relation, but we control the function only close to the threshold and the behavior at infinity generates impractical subtraction constants. The solution can be instead easily guessed~\cite{Correia:2020xtr}.

Let us start with the case \textbf{$d$ odd and integer}, $d \in (2\mathbb{Z}+1)$. In this case, $\alpha$ is an integer and we look for a function, $1/f_0(s)$ whose discontinuity is given by a polynomial. A trivial solution is the logarithm, whose discontinuity is $\im \log(4-s)=-\pi$, for which we get:
\begin{equation}
\label{eq:inverse-f0-odd}
2\frac{1}{f_0(s) } =  \frac{(4-s)^{\al}}{\sqrt{s}} \pi \log(4-s)+b_0(s),\qquad s<4 \,,
\end{equation}
where $b_0(s)$ is any analytic function near $s=4$, admitting a Taylor expansion $b_0(s)= \sum_{n=0}^\infty b_{0,n} (s-4)^n$. This logarithm was explicitly computed in $\phi^4$ in $d=3$ in~\cite{Bros:1998tt}, as mentioned in the main text.
Note that the scale of the logarithm is irrelevant here and can always be absorbed in the Taylor coefficient $b_{0,\alpha}$.
This problem matters below when we cook up the threshold term for the ansatz, as the logarithm creates a pole, which should be removed, see the discussion below eq.~\eqref{eq:thDelta}.

For $d$ \textbf{not odd and real}, $d\in \mathbb{R} \backslash (2\mathbb{Z}+1)$, the most generic form of the solution is likewise found to be:
\begin{equation}
\label{eq:inverse-f0-generic}
2\frac{1}{f_0(s) } = \frac1{\sin(\pi\al)} \frac{(4-s)^{\al}}{\sqrt{s}}+b_0(s),\qquad s<4 \,,
\end{equation}
where again $b_0(s)$ is an analytic function near $s=4$.
The phase factor $\sin(\pi\al)$ comes from the multivaluedness of the phase-space factor for $d$ non-integer, and it can be checked explicitly that this function satisfies unitarity near the two-particle threshold in non-integer $d$, noting that we get:
\begin{equation}
    f_0(s) = \sin(\pi\al)\frac{2\sqrt{s}}{(4-s)^\al},\quad \Re(f_0(s)) = \cos(\pi\al)\sin(\pi\al)\frac{2\sqrt{s}}{(4-s)^\al}\,,
    \quad
    \Im(f_0(s))= \sin(\pi\al)^2\frac{2\sqrt{s}}{(4-s)^\al}\,,
\end{equation}
for $s=|s|+i 0^+$, $|s|>4$.

\subsection{Engineering a divergent term for $f_0(s)$}
From these solutions for $1/f_0(s)$, we get immediately the near threshold expansion of $f_0(s)$. 

Starting with $d$ odd and $d\geq3$, $\alpha$ is a positive integer, 
and we can invert \eqref{eq:inverse-f0-odd} near $s=4$ (we replace $\sqrt{s}$ in the phase-space factor by $2$):
\begin{equation}
f_0(s) =  \frac{2}{(4-s)^{\alpha}} \frac{1}{\pi \log(4-s)+ \frac12\sum_{n=0}^\infty b_{0,n} (s-4)^{n-\alpha}}\,.
\end{equation}
The singular/regular behavior near the threshold is dictated by the vanishing of first few Taylor coefficients in $b_0(s)$. If the first $\alpha$ coefficients vanish, we obtain a maximally divergent threshold, for which we have
\begin{equation}
\label{eq:taylor-odd}
f_0(s) =  \frac{2}{(4-s)^{\alpha}} \frac{1}{\pi \log(4-s)}+\text{subleading}\,.
\end{equation}
The subleading terms can be extracted by 
consistently Taylor expanding in powers of $(s-4)$. For instance, if only the first $\alpha-1$ coefficients vanish, the leading term near threshold is given by
\begin{equation}
f_0(s) =  \frac{2}{(s-4)^{\alpha}} \frac{\frac{2}{b_{0,\alpha-1}}(s-4)}{\pi \log(4-s)}+\text{subleading}\,,
\end{equation}
and so on.

For instance, for $d=5, \alpha=1$ and near $s=4$, the maximally divergent $f_0(s)$ behaves like
\begin{multline}
\label{eq:5d-ex}
d=5\qquad f_0(s) =  \frac{2}{(s-4)} \frac{1}{\pi \log(4-s)+ \frac12 b_{0,1}+ \frac12 b_{0,2}(s-4) +\dots } =  \\ \frac{2}{\pi(s-4)\log(4-s)}
\bigg[1-\frac 12 b_{0,1}\frac{1}{\pi \log(4-s)}+\left(\frac 12 b_{0,1}\frac{1}{\pi \log(4-s)}\right)^2+\dots\\
-\frac 12 b_{0,2}\frac{s-4}{\pi \log(4-s)}+\left(\frac 12 b_{0,2}\frac{s-4}{\pi \log(4-s)}\right)^2+\dots
\bigg]\,.
\end{multline}
As it happens, infinitely many divergent logarithmic subleading threshold corrections actually need to be included, all coming from the $b_{0,1}$ term. In this example, the $b_{0,2}$ corrections are regular. %
Note that the leading order coefficient in \eqref{eq:taylor-odd}, corresponding to the $1$ above, is entirely fixed by elastic unitarity. The subleading contributions are not and depend on the free coefficient $b_{0,1}$.

Let us now turn to the generic \textbf{non-odd and real} $d$ case is simpler and finitely many terms contribute to the divergent pieces. Starting from the generic solution
\begin{equation}
\label{eq:taylor-non-odd}
f_0(s) =  \frac{2}{\sin(\pi \alpha)(4-s)^{\alpha}} \frac{1}{1+ \frac12\sum_{n=0}^\infty b_{0,n} (s-4)^{n-\alpha}}\,,
\end{equation}
the leading order divergence occurs again when $b_{0,0}=\dots=b_{0,\lfloor \alpha \rfloor}=0$, so that
\begin{equation}
f_0(s) =  \frac{2}{\sin(\pi \alpha)(4-s)^{\alpha}}
+\text{subleading}\,.
\end{equation}
The subleading terms are immediate to work out again by means of a Taylor expansion of the general solution. 
For instance, in $d=6$, $\alpha=3/2$ and the most divergent solution has the following leading and subleading threshold divergent pieces:
\begin{multline}
\label{eq:6d-ex}
d=6\,,\qquad f_0(s) =  \frac{2}{(s-4)^{3/2}} \frac{1}{1+ \frac12 b_{0,2}\sqrt{s-4}+ \frac12 b_{0,3}(s-4)^{3/2} +\dots } =  \\\frac{2}{(s-4)^{3/2}}
\bigg[1-\frac 12 b_{0,2}\sqrt{s-4}+\left(\frac 12 b_{0,2}\sqrt{s-4}\right)^2
\bigg]+\text{regular}\,.
\end{multline}
We provide an explicit form of these corrections below in eq.~\eqref{eq:th-subth-all} for the relevant dimensions.

For $d$ half-integer, new powers multiple of $1/4$ appear. For generic rational $d$, a trivially determined but possibly very large set of other powers enter. The practical consequences of this fact, easy to work out, make it a priori easier to approach integer dimensions from above rather than form below.

\subsection{Engineering a divergent term for the amplitude, $A_\text{th}(s,t)$}
\label{app:art-TH}
Having worked out in details the structure of $f_0(s)$, we need to produce an ansatz for the scattering amplitude that gives rise to these divergent structures while maintaining crossing symmetry.
In four dimensions, as was noted early in~\cite{Paulos:2017fhb}, the structure is simple and one writes:
\begin{equation}
   d=4\,,\qquad A_{\rm th}(s,t) = \frac{64\pi}{\sqrt{s-4}}+\frac{64\pi}{\sqrt{t-4}}+\frac{64\pi}{\sqrt{u-4}}\,.
\end{equation}
It develops an imaginary part beyond two-particle threshold in every channel and it has no double discontinuity, hence contributes only to spin-zero, as expected. The coefficient $64\pi$ comes from the partial wave expansion and is fixed by elastic unitarity~\eqref{eq:unitarity_el} as explained above.

In \textbf{non-odd and real} higher dimensions, the idea can be implemented similarly. Non-analyticities in the threshold expansion of \ref{eq:taylor-non-odd} are all localized to $s > 4$, and we can safely combine them with the crossed pieces obtained by simply exchanging $s \to t$ and $s \to u$, giving rise to  the explicit formula below in eq.~\eqref{eq:ath}.

However, for \textbf{odd integer} dimensions, there arises a technical problem in constructing the crossing symmetric ansatz. The threshold expansion for $f_0(s)$ in \eqref{eq:taylor-odd} contains an overall $\log((4-s)/4)^{-1}$ which puts a simple pole at $s=0$. We cannot lift it up to the amplitude level, since the simple pole (and its copies under crossing $t=0$ and $u=0$) bring spurious singularities violating our initial analyticity assumptions. Instead we need to devise a way to get rid of the unwanted pole. To this purpose, we dress it the threshold term with the squared wavelet $\rho(s,8)^2$ as follows
\begin{equation}
\label{eq:log-shorthand}
\frac{1}{\log_s}\equiv\frac{\pi}{\log\left(\frac{s-4}{4}\right)}\left(\frac{2-\sqrt{4-s}}{2+\sqrt{4-s}}\right)^2 \,,
\end{equation}
where we use the shorthand notation $\frac{1}{\log_s}$ below.
The wavelet dressing cancels the pole at $s=0$ an (i) does not cause a growth at infinity, contrary to any power of $s$ would have done, (ii) does not change the overall analytic structure, since it only has the same type of branch cuts $\sqrt{4-s}$ as the rest of the terms in the ansatz. Engineering this term was key to obtain well-converged and reliable bounds in odd $d$. 

In the end, the threshold improvement terms to be supplied in the ansatz \eqref{eq:ansatz} are given by
\begin{equation}
\label{eq:ath}
A_{\text{th}}(s,t) = n_d \, \sum_{n \in \mathcal{I}} \,
\beta_n \left[
\left({1-\frac{s}{4}}\right)^{-\tfrac{d-3}{2}-n} \Delta_n^{(d)}(s)  
+(s\to t)+(s\to u)
\right]\,,
 \end{equation}
where $n_d$ is given in \eqref{eq:nd}, and
\begin{equation}
\Delta_n^{(d)}(s)=\sin\left(\pi\tfrac{d-3}{2}\right)\times
\begin{cases}
\cfrac{\pi }{\log(\tfrac{4-s}{4})} \, \rho(s,8)^2 & d \text{ odd} , \quad d \in (2\mathbb{Z}+1)\\
1 & d \text{ non-odd}, \quad d \in \mathbb{R} \backslash (2\mathbb{Z}+1)
\label{eq:thDelta}
\end{cases}\,.
\end{equation}
The summing index $n$ runs over the set $\mathcal{I}$ of leading ($n=0$) and subleading terms in the threshold expansion of $f_0(s)$ as explained in the previous subsection. 

For odd d, $\mathcal{I}=\{0,1,2 \dots (d-3)/2\}$. However, in the odd case, we observed that this procedure was not generating sufficiently many terms to give enough numerical freedom to threshold unitarization, so we doubled the number of terms, by including extra $\log$ squared terms.

For non-odd $d$,
$\mathcal{I}$ contains the rational numbers starting from $3-d+\lceil(d-3)/2\rceil$ to 0 by steps of $\lceil(d-3)/2\rceil-(d-3)/2$. 
Refer to the Table at the end of the section for a full list of generated terms.

The real coefficients $\beta_n$ become free parameters of the ansatz, analogous to $\alpha_{ijkl}$ in \eqref{eq:ansatz}. Note that elastic unitarity restricts the coefficient of the leading singularity to have only two values, either zero or a fixed non-zero constant, which was a good diagnostic for us to assess the numerical convergence near the threshold. This also motivates the $n_d$ normalization in front, since it is chosen so that $\beta_0 \in \{0,1\}$. The remaining coefficients $\beta_{n>0}$ are not restricted and determined dynamically by SDPB to unitarize the partial waves $<J_\text{max}$.


Finally, here we explicitly give the full list of threshold terms in all dimensions studied in this letter:
\begin{equation}
\resizebox{0.93\textwidth}{!}{$
\begin{aligned}
& d=3:   && \frac{\beta_0}{\log_s} + 
         \frac{\beta'_0}{\log_s^2}  \\
& d=3.5: && \frac{\beta_0}{\sigma_s^{1/4}} \\
& d=4:   && \frac{\beta_0}{\sigma_s^{1/2}} \\
& d=4.5: && \frac{\beta_0}{\sigma_s^{3/4}} + 
         \frac{\beta_1}{\sigma_s^{1/2}} + 
         \frac{\beta_2}{\sigma_s^{1/4}}\\
& d=5:   && \frac{\beta_0}{\sigma_s\log_s}+
         \frac{\beta_1}{\log_s}+
         \frac{\beta'_0}{\sigma_s\log_s^2}+
         \frac{\beta'_1}{\log_s^2}\\
& d=5.5: && \frac{\beta_0}{\sigma_s^{5/4}}+
         \frac{\beta_1}{\sigma_s^{1/2}}\\
& d=6:   && \frac{\beta_0}{\sigma_s^{3/2}}+
         \frac{\beta_1}{\sigma_s}+
         \frac{\beta_2}{\sigma_s^{1/2}}\\
& d=6.5: && \frac{\beta_0}{\sigma_s^{7/2}}+
         \frac{\beta_1}{\sigma_s^{3/2}}+
         \frac{\beta_2}{\sigma_s^{5/4}}+
         \frac{\beta_3}{\sigma_s}+
         \frac{\beta_4}{\sigma_s^{3/4}}+
         \frac{\beta_5}{\sigma_s^{1/2}}+
         \frac{\beta_6}{\sigma_s^{1/4}}\\
& d=7:   && \frac{\beta_0}{\sigma_s^{2}\log_s}+
         \frac{\beta_1}{\sigma_s\log_s}+
         \frac{\beta_2}{\log_s}+
         \frac{\beta'_0}{\sigma_s^{2}\log_s^2}+
         \frac{\beta'_1}{\sigma_s\log_s^2}+
         \frac{\beta'_2}{\log_s^2}\\
& d=7.5: && \frac{\beta_0}{\sigma_s^{9/4}}+
         \frac{\beta_1}{\sigma_s^{3/2}}+
         \frac{\beta_2}{\sigma_s^{3/4}}\\
& d=8:   && \frac{\beta_0}{\sigma_s^{5/2}}+
         \frac{\beta_1}{\sigma_s^{2}}+
         \frac{\beta_2}{\sigma_s^{3/2}}+
         \frac{\beta_3}{\sigma_s}+
         \frac{\beta_4}{\sigma_s^{1/2}}\\
& d=8.5: && \frac{\beta_0}{\sigma_s^{11/4}}+
         \frac{\beta_1}{\sigma_s^{5/2}}+
         \frac{\beta_2}{\sigma_s^{1/4}}+
         \frac{\beta_3}{\sigma_s^{9/4}}+
         \frac{\beta_4}{\sigma_s^{2}}+
         \frac{\beta_5}{\sigma_s^{7/4}}+
         \frac{\beta_6}{\sigma_s^{3/2}}+
         \frac{\beta_7}{\sigma_s^{5/4}}+
         \frac{\beta_8}{\sigma_s}+
         \frac{\beta_9}{\sigma_s^{3/4}}+
         \frac{\beta_{10}}{\sigma_s^{1/2}}\\
& d=9:   && \frac{\beta_0}{\sigma_s^{3}\log_s}+
         \frac{\beta_1}{\sigma_s^{2}\log_s}+
         \frac{\beta_2}{\sigma_s\log_s}+
         \frac{\beta_3}{\log_s}+
         \frac{\beta'_0}{\sigma_s^{3}\log_s^2}+
         \frac{\beta'_1}{\sigma_s^{2}\log_s^2}+
         \frac{\beta'_2}{\sigma_s\log_s^2}+
         \frac{\beta'_3}{\log_s^2}\\
& d=9.5: && \frac{\beta_0}{\sigma_s^{13/4}}+
         \frac{\beta_1}{\sigma_s^{5/2}}+
         \frac{\beta_2}{\sigma_s^{7/4}}+
         \frac{\beta_3}{\sigma_s}+
         \frac{\beta_4}{\sigma_s^{1/4}}\\
& d=10:  && \frac{\beta_0}{\sigma_s^{7/2}}+
         \frac{\beta_1}{\sigma_s^{3}}+
         \frac{\beta_2}{\sigma_s^{5/2}}+
         \frac{\beta_3}{\sigma_s^{2}}+
         \frac{\beta_4}{\sigma_s^{3/2}}+
         \frac{\beta_5}{\sigma_s}+
         \frac{\beta_6}{\sigma_s^{1/2}}\\
& d=10.5:&& \frac{\beta_0}{\sigma_s^{15/4}}+
         \frac{\beta_1}{\sigma_s^{7/2}}+
         \frac{\beta_2}{\sigma_s^{5/4}}+
         \frac{\beta_3}{\sigma_s}+
         \frac{\beta_4}{\sigma_s^{3/4}}+
         \frac{\beta_5}{\sigma_s^{1/2}}+
         \frac{\beta_6}{\sigma_s^{1/4}}+
         \frac{\beta_7}{\sigma_s^{13/4}}+
         \frac{\beta_8}{\sigma_s^{3}}+
         \frac{\beta_9}{\sigma_s^{11/4}}+
         \frac{\beta_{10}}{\sigma_s^{5/2}}+
         \frac{\beta_{11}}{\sigma_s^{9/4}}+
         \frac{\beta_{12}}{\sigma_s^{2}}+
         \frac{\beta_{13}}{\sigma_s^{7/4}}+
         \frac{\beta_{14}}{\sigma_s^{3/2}}\\
& d=11:  && \frac{\beta_0}{\sigma_s^{4}\log_s}+
         \frac{\beta_1}{\sigma_s^{3}\log_s}+
         \frac{\beta_2}{\sigma_s^{2}\log_s}+
         \frac{\beta_3}{\sigma_s\log_s}+
         \frac{\beta_4}{\log_s}+
         \frac{\beta'_0}{\sigma_s^{4}\log_s^2}+
         \frac{\beta'_1}{\sigma_s^{3}\log_s^2}+
         \frac{\beta'_2}{\sigma_s^{2}\log_s^2}+
         \frac{\beta'_3}{\sigma_s\log_s^2}+
         \frac{\beta'_4}{\log_s^2}\\
& d=11.5:&& \frac{\beta_0}{\sigma_s^{17/4}}+
         \frac{\beta_1}{\sigma_s^{7/2}}+
         \frac{\beta_2}{\sigma_s^{11/4}}+
         \frac{\beta_3}{\sigma_s^{2}}+
         \frac{\beta_4}{\sigma_s^{5/4}}+
         \frac{\beta_5}{\sigma_s^{1/2}}\\
& d=12:  && \frac{\beta_0}{\sigma_s^{9/2}}+
         \frac{\beta_1}{\sigma_s^{4}}+
         \frac{\beta_2}{\sigma_s^{7/2}}+
         \frac{\beta_3}{\sigma_s^{3}}+
         \frac{\beta_4}{\sigma_s^{5/2}}+
         \frac{\beta_5}{\sigma_s^{2}}+
         \frac{\beta_6}{\sigma_s^{3/2}}+
         \frac{\beta_7}{\sigma_s}+
         \frac{\beta_8}{\sigma_s^{1/2}}
\end{aligned}
$}
\label{eq:th-subth-all}
\end{equation}
where $\log_s$ is a shorthand given in \eqref{eq:log-shorthand}, and $\sigma_s=s{-}4$, and the $n_d$ normalization and other signs and constants has been absorbed in the $\beta$'s for simplicity.

\section{Numerical bootstrap setup}
\label{app:primal_bootstrap}

We use the following primal ansatz for the amplitude in arbitrary $d$ spacetime dimensions
\begin{equation}
A(s,t) = A_\text{th}(s,t)+\sum^{N_\text{max}}_{i,j=1} \sum^1_{k,l=0} \,\alpha_{ijkl}
\,\mathbf{P}_{ijkl}(s,t)\,,
\label{eq:ansatz}
\end{equation}
where $\alpha_{ijkl}$ are real-valued free parameters and $A_\text{th}$ is constructed in Appendix~\ref{app:threshold}. The wavelet-inspired multi-foliated basis~\cite{EliasMiro:2022xaa,Guerrieri:2024jkn} is given by
\begin{equation}
\mathbf{P}_{ijkl}(s,t)=\frac{n_d}{6} \, \left(\rho(s,\sigma_i)^k \cdot \rho(t,\sigma_j)^l + s \leftrightarrow t \leftrightarrow u \right)\,\delta_{ijkl}\,, 
\label{eq:Pansatz}
\end{equation}
with $n_d$ given in \eqref{eq:nd}. The basis $\mathbf{P}_{ijkl}(s,t)$ is given as a crossing-symmetric sum over double products of the conformal map
\begin{equation}
\rho(s,\sigma)=\frac{\sqrt{\sigma-4}-\sqrt{4-s}}{\sqrt{\sigma-4}+\sqrt{4-s}}\,,
\label{eq:confmap}
\end{equation}
and where the first two indices
$i,j \in \{1 \dots N_{\mathrm{max}}\}$ counts through the wavelet centers $\sigma_i$ (defined below) and the last two indices through their powers $k \in \pi_i=\{0,1\}$, $l \in \pi_j=\{0,1\}$.

Finally, we introduced the following condition to avoid the double counting
%
\begin{equation}
\delta_{ijkl}=\begin{cases}
    1, \text{ if } Q \text{ true} \\ 0, \text{ if not }
\end{cases}
\end{equation}
with
$Q=
\sigma_i\leq\sigma_j \, \land \, 
k \leq l \, \land \,
(k\neq0 \lor i\leq 1) \, \land \,
(l\neq0 \lor j\leq 1) \, \land \,
k\leq \min(\pi_i,\pi_j) \, \land \,
l\leq \max(\pi_i,\pi_j)\,.
$
Although the condition is intricate, it describes precisely how to avoid double-counting.

\subsection{Samplings}
We require two discrete grids to implement the numerics: One we call $\{ s_n \}$ is to impose unitarity condition \eqref{eq:unitarity2}, and another one we call $\{ \sigma_n \}$ to distribute the wavelet centers in the ansatz \eqref{eq:ansatz}. Together with the spin cutoff for the unitarized (even) partial wave coefficients, they respectively introduce the truncation parameters $(N_\text{sgrid}, N_\text{max}, J_\text{max})$ in our setup to be sent to infinity. 

Our typical choices for the constraints are $N_\text{sgrid}=300$ and $J_\text{max}=16$. We checked that a more dense unitarity grid, or reducing the spin cutoff to $J_\text{max}=14$, affects our results only up to the second digit after the decimal point in the worst case across all $d$.
The dependence of our results and convergence on the ansatz size $N_\text{max}$ is discussed later below. Next, we lay out how we construct aforementioned grids explicitly.

Let us start from a Chebyshev grid sampling the upper half circle
\begin{equation}
\phi_n =\frac{\pi}{2}\left(1+\cos\frac{n\pi}{N_\text{sgrid}+1}\right) \in (0,\pi)\,, \qquad n \in \{1 \dots N_\text{sgrid}\}\,.
\end{equation}

We can apply the conformal map on $\{\phi_n\}$ to sample the center-of-mass energies from two-particle threshold up to arbitrarily high energies, thus generating the unitarity grid
\begin{equation}
s_n=\rho^{-1}(e^{i\phi_n},\sigma_0) \in (4,\infty)\,,
\end{equation}
where $\rho^{-1}$ is the inverse of the conformal map \eqref{eq:confmap} and $\sigma_0=20/3$ stands for the mother wavelet center that solves $\rho(4/3,\sigma_0)=0$. For our choice $N_\text{sgrid}=300$, the unitarity grid $\{s_n\}$ has an extent $s-4 \in [10^{-9},10^9]$.

We will use the unitarity grid to generate cumulatively the wavelet centers grid. Consider the ordered set $(s_n)$ of grid points. We refer to its $i$-th element by $(s_n)_i$ and its size by $|s_n|$.
Then, we construct
$$\Sigma_m = \bigcup\limits_{i=1}^{m} s_i$$
and choose a critical $m(n)$ for a given $n$ such that $|\Sigma_{m(n)}| \geq n \geq |\Sigma_{m(n)-1}|$. Then,
\begin{equation}
\sigma_1=s_1 \quad , \quad
\sigma_n=(\Sigma_{m(n)})_{n} \quad , \quad n \in \{1 \dots N_\mathrm{max}\}
\label{eq:sigmagrid}
\end{equation}
Note that $|\sigma_n|=n$. For reference, we provide a table below for the wavelet centers that are shot at each order:
\begin{equation}
\begin{array}{|c|c|c|c|c|c|c|c|c|c|c|}
\hline
N_{\mathrm{max}} & 1 & 2 & 3 & 4 & 5 & 6 & 7 & 8 & 9 & 10 \\ \hline
\sigma & 6.667 & 4.458 & 19.54 & 4.146 & 52.62 & 4.061 & 4.970 & 11.33 & 120.7 & 4.030 \\ 
\hline \hline
N_{\mathrm{max}} & 11 & 12 & 13 & 14 & 15 & 16 & 17 & 18 & 19 & 20 \\ \hline
\sigma & 243.1 & 4.016 & 4.247 & 5.306 & 9.444 & 32.73 & 443.0 & 4.010 & 4.740 & 13.61 \\
\hline \hline
N_{\mathrm{max}} & 21 & 22 & 23 & 24 & 25 & 26 & 27 & 28 & 29 & 30 \\ \hline
\sigma & 748.3 & 4.006 & 4.092 & 5.535 & 8.633 & 81.21 & 1191. & 4.004 & 4.300 & 27.68 \\ \hline
\end{array}
\end{equation}
The number of the terms in the ansatz \eqref{eq:ansatz} (not including threshold terms) is given by 
\begin{equation}
\frac{1}{2}(N_\text{max}+1)(N_\text{max}+2) \propto N_\text{max}^2\,.
\end{equation}
For instance, it corresponds to 231 independent parameters when $N_{\mathrm{max}}=20$, and 465 when $N_{\max}=29$.

\subsection{Optimization problem}
\label{app:primal_opt_prob}
\vspace{-0.2cm}

The primal ansatz \eqref{eq:ansatz} automatically satisfies maximal analyticity and crossing symmetry.  Unitarity \eqref{eq:unitarity2} as the only remaining condition will be solved for using semi-definite optimization tools. We can recast it in terms of the the following $2\times2$ matrix
\begin{equation}
U_J(s)=\begin{pmatrix}
1-\frac{1}{2}\phi_d^2(s)\operatorname{Im} f_J(s) &
\phi_d(s)\operatorname{Re}f_J(s)  \\
\phi_d(s)\operatorname{Re}f_J(s) &
2\operatorname{Im}f_J(s)
\end{pmatrix}\, \succeq 0 
\quad , \quad J \in \{0,2 \dots J_\text{max}\} \quad , \quad s \in \{s_n\} \, .
\label{eq:SDP-PWU}
\end{equation}
For our typical choices, it gives rise to $N_{\rm sgrid}\times (1+J_{\max}/2) = 300 \times 9 =2700$ non-linear constraints on $\alpha_{ijkl}$ and $\beta_i$, each one containing $\sim N_{\max}^4$ terms. 

We also add \emph{subtracted positivity constraints}~\cite{Guerrieri:2021ivu,EliasMiro:2022xaa} ,
\begin{equation}
\text{SPC}(s,t) = \im \, A(s,t)-
\sum_{J=0,\atop J~\text{even}}^{J_\mathrm{max}} n^{(d)}_J
P^{(d)}_J \left(1+\frac{2t}{s-4}\right)
\, \im \, f_J(s) \geq 0 \quad , 
\label{eq:SPC}
\end{equation}
sampled over the grid $\{ s_n \}$ and we define a smaller grid $\{t_n\}$ of size $N_{\rm tgrid}=10$, with finite momentum transfer values
\begin{equation}
t_n \in \{0.27, 1, 2, 3, 3.73, 3.99994, 3.99996, \
3.99998, 3.99999, 4.00000\} \,,
\end{equation}
giving $N_{\rm sgrid}\times (1+J_{\max}/2) \times N_{\rm tgrid}= 300 \times 9 \times 10=27000$ linear constraints of $\sim N_{\max}^2$ terms. These conditions imposes positivity on infinitely many partial waves above the cutoff $J > J_\text{max}$, improving convergence drastically. 

Finally, we formulate the following optimization problem
\begin{equation}
\begin{aligned}
\text{optimize} \quad & c_n \\
\text{over} \quad & \{ \alpha_{ijkl}, \beta_i \} \\
\text{subject to} \quad & \eqref{eq:SDP-PWU}, \eqref{eq:SPC} \, .
\end{aligned}
\end{equation}
At the extremal solution, we can plug in the coefficients $\{ \alpha_{ijkl}, \beta_i \}$ into \eqref{eq:ansatz}, and obtain the partial amplitudes $S_J(s)$.

\vspace{-0.2cm}
\subsection{Convergence and extrapolation as a function of the ansatz size $N_{\max}$.}
\label{app:convergence}

\begin{figure*}[h!]
    \centering
    \begin{subfigure}{0.45\textwidth}
        \centering
        \includegraphics[width=0.9\linewidth]{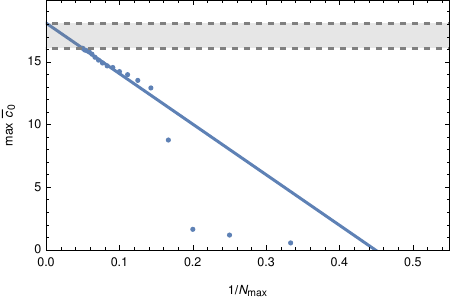}
        \vspace{-0.1cm}
        \caption{$\max c_0$ as a function of $1/N_{\max}$ in $d=8$ with simple linear fit over the last 10 points.}
        \label{fig:nmaxfit1}
    \end{subfigure}
    \begin{subfigure}{0.45\textwidth}
        \centering
        \includegraphics[width=0.9\linewidth]{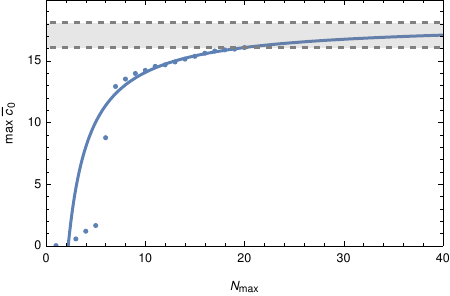}
        \vspace{-0.1cm}
        \caption{$\max c_0$ as a function of $N_{\max}$ in $d=8$, with same linear fit reproduced in inverse coordinates.}
        \label{fig:nmaxfit2}
    \end{subfigure}
    \vspace{-0.2cm}
    \caption{Example of our extrapolation method in $d=8$. Lower dashed line is the best value obtained at finite truncation, here $\max c_0(N_{\max}{=}20)=16.1$, while upper dashed line is the extrapolation of the simple linear fit on last 10 points, here $\max c_0(N_{\max}\rightarrow\infty)=18.1$. We consider these two values respectively as lower and upper values on our best estimate, which we hence define here as $\max c_0=17.1\pm1$ in $d=8$.}
    \label{fig:extrapolation}
\end{figure*}

\vspace{-0.2cm}
In Figure~\ref{fig:extrapolation}, we explain our extrapolation method in $N_{\max}$, and how error bars in Figure~\ref{fig:fourpanel} are produced by  providing a concrete example in $d=8$. Of course, this extrapolation method is not unique. We choose this one because of its simplicity and its reliability over the wide span of dimensions we study in this paper. Final results are barely affected if we choose to fit over 7, 8, 9 points instead of 10. Its also the best method we could find so that the produced error bar reflect our actual ignorance on the bound, rather than taking the error of the chosen fit method itself. 
\vspace{-0.15cm}

\newpage

\section{IR subtractions, positivity of Wilson coefficients, and the loss thereof}
\label{app:ir_sub}

The presence of singular threshold behavior in $d\geq 5$ has important consequences for dispersive representations of the amplitude and for the positivity properties of certain low-energy observables. For instance, in $d=4$, the four-derivative quartic interaction coefficient $c_2$ admits a standard dispersive representation as a positive moment of the imaginary part of the amplitude (that we derive below in~\eqref{eq:c2-pos}), leading to the familiar positivity bound $c_2\geq 0$~\cite{Pham:1985cr,Adams:2006sv}. This argument relies, in particular, on the integrability of the imaginary part near the two-particle threshold. 

\begin{figure}[h!]
    \centering
    \includegraphics[scale=1]{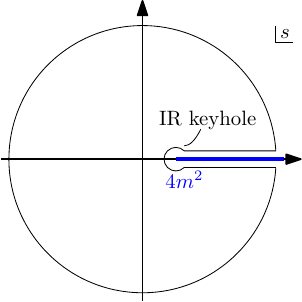}
    \caption{An typical keyhole contour in a dispersion relation, regulating the infrared divergence at $s=4m^2$.}
    \label{fig:ir_keyhole}
\end{figure}

In higher dimensions, the singular behavior allowed by elastic unitarity renders the kernel of twice-subtracted fixed-$t$ dispersion relation unintegrable at threshold. Of course, result of the dispersive Cauchy integral must still be finite, when one regulates the divergence with a keyhole integral contour (see Figure~\ref{fig:ir_keyhole}), for instance. Such regularizations are also encountered when looking at large $J$ limits of partial waves and threshold expansion of amplitude's double-discontinuity~\cite{Correia:2020xtr}.

Alternatively, we can introduce additional infrared subtractions, analogous to the ultraviolet subtractions required when amplitudes grow at high energy. The only difference is that now we add extra \textit{zeros} to soften the IR, namely to absorb the power-like divergences that sit at the $s-$ and $u-$channel thresholds branch cuts $s=4$ and $s=-t$ at fixed momentum transfer $t$. With this motivation, we consider a dispersion integral of the following form
\begin{equation}
    A(s,t) = \frac{1}{2i\pi} \oint_s ds' \frac{A(s',t)}{s'-s}
    \left[ \frac{s-s_0}{s'-s_0} \, \frac{s-4+s_0+t}{s'-4+s_0+t} \right]^{\frac{D-2}{2}}
    \left[ \frac{s'-4}{s-4} \, \frac{s'+t}{s+t} \right]^{\frac{D-4}{2}}\,,
    \label{eq:ir_cauchy}
\end{equation}
where $D/2=\lceil d/2 \rceil$. The kernel vanishes quadratically when $|s'| \to \infty$, as usual for a twice-ultraviolet-subtracted dispersion relation. Deforming the contour and picking up singularities in the complex $s'$-plane in the standard fashion yields new \emph{IR-subtracted dispersion relations} at fixed-$t$. They express the amplitude in terms of a convolution integral of its imaginary part and subtraction terms which include the first $D/2{-}1$ derivatives of the amplitude at $s=s_0$.  For convenience, we fix $s_0=t=4/3$ from now on. %

In these new dispersion relations, certain low-energy coefficients—including $c_2$ for $d\geq 5$—become independent subtraction constants in the dispersive representation. They are no longer determined by a positive spectral integral, and there is therefore no general reason for them to satisfy positivity constraints. This loss of positivity is an immediate consequence of the allowed threshold singularities and does not rely on any additional dynamical assumptions.

Generically, the positivity of 
\begin{equation}
    c_{2n} = \frac{1}{2^{n}}\frac{1}{(2n)!}\partial^{2n}_{s} A(s,\tfrac{4}{3})|_{s=\frac{4}{3}}\,,
\end{equation}
is lost when $2n \leq d-4$ for an amplitude with singular threshold behavior. Of course, if the threshold behavior is instead regular, then positivity is recovered. Note that the $2^n(2n)!$ normalization is chosen so that the low energy expansion \eqref{eq:wilsons} include the even terms: $A(s,t) \subset \sum_{n=0}^\infty c_{2n} (\bar s^2+\bar t^2+\bar u^2)^n$.

We note that it is still possible to construct new positive-definite combinations of low-energy coefficients, by applying sufficiently many derivatives to the IR-subtracted dispersion relation~\eqref{eq:ir_cauchy}. 
For instance, applying $\partial_s^4$ on both sides of~\eqref{eq:ir_cauchy} in $d=6$, we obtain (note the softening factor $(s'-4)$ in front of $\im A$)
\begin{equation}
\label{eq:c4-pos}
    c^+_{4}\equiv c_2 - \tfrac{128}{9}c_4 = \int_{4}^\infty \frac{ds'}{\pi} \frac{(s'{+}\tfrac{4}{3})}{(s'{-}\tfrac{4}{3})^5} \, (s'{-}4) \im A(s',\tfrac{4}{3})\geq 0.
\end{equation}
Similarly, applying $\partial_s^2$ in $d=4$ gives rise to the well-known
\begin{equation}
\label{eq:c2-pos}
    c^+_{2}\equiv c_2 =  \int_4^\infty \frac{ds'}{\pi} \, \frac{\im A(s',\tfrac{4}{3})}{(s'{-}\tfrac{4}{3})^3} \, \geq 0 \, .
\end{equation}

Notably, for each dimension $d$, the procedure of iterating $\partial_s$ yields an infinite family of positive quantities involving linear combinations of the coefficients $c_{2n}$, of the following form
\begin{equation}
c^+_{2n} = c_{2n} + \sum^{D/2-1}_{k=2 
} F_k(d)\,c_k  \geq 0\,,
\label{eq:pos_c2n}
\end{equation}
with $F_k(d)\in \mathbb{Q}$ for fixed $d$ and $k$.

Up to $d \leq 5$, \eqref{eq:pos_c2n} recovers the known infinite tower $c^+_{2n}=c_{2n}\geq0$.
For $d>5$ and beyond, a new tower emerges each time an odd dimension is reached and a finite number of $c_2, c_4 \dots$ become subtraction constants, while the positivity on the individual $\{c_{2n}\}$ is compromised. Refer to Table~\ref{tab:positivity} for various positive-definite linear combinations with respect to $d$. Table~\ref{tab:positivity} also shows a quite interesting fact that, in $d \leq 7$, if $c_2$ is negative, then all of the other $c_{2n}$'s have to be negative as well, meaning that $\{c_{2n}\}$ are either all positive or all negative!

%
%

\begin{table}[h!]
\begin{equation*}
\renewcommand{\arraystretch}{1.5}
\begin{array}{|c|c|c|c|c|c|c|}
\hline
& 3{<}d{\leq}5 & d\leq7 & d\leq9 & d\leq11 & \cdots & d\leq(2\lambda+1) \\
\hline
c^+_{2}
& c_{2}
& ✗
& ✗
& ✗
& \dots
& ✗
\\
\hline
c^+_{4}
& c_{4}
& c_{2}{-}\tfrac{128}{9}  c_{4}
& ✗
& ✗
& \dots
& ✗
\\
\hline
c^+_{6}
& c_{6}
& c_{2}{-}\tfrac{16384}{81}  c_{6}
& c_{2}{-}\tfrac{256}{9} c_{4}{+}\tfrac{16384}{81}  c_{6}
& ✗
& \dots
& ✗
\\
\hline
c^+_{8}
& c_{8}
& c_{2}{-}\tfrac{2097152}{729} c_{8}
& c_{2}{-}\tfrac{64}{3} c_{4}{+}\tfrac{1048576}{729} c_{8}
& c_{2}{-}\tfrac{128}{3} c_{4}{+}\tfrac{16384}{27} c_{6}{-}\tfrac{2097152}{729} c_{8}
& \dots
& ✗
\\
\hline
c^+_{10}
& c_{10}
& c_{2}{-}\tfrac{268435456}{6561} c_{10}
& c_{2}{-}\tfrac{512}{27} c_{4}{+}\tfrac{268435456}{19683} c_{10}
& c_{2}{-}\tfrac{1024}{27} c_{4}{+}\tfrac{32768}{81} c_{6}{-}\tfrac{268435456}{19683} c_{10}
& \dots
& ✗
\\
\hline
\vdots
& \vdots
& \vdots
& \vdots
& \vdots
& \ddots
& \vdots
\\
\hline
c^+_{2n}
& c_{2n}
& c_{2}{-}\# c_{2n}
& c_{2}{-}\# c_{4}{+}\# c_{2n}
& c_{2}{-}\# c_{4}{+}\# c_{6}{-}\# c_{2n}
& \dots
& (-1)^\lambda c_{2n} \! - \!\!\!\!\!\! \displaystyle \sum_{j=1}^{\lfloor \lambda-3/2 \rfloor}\!\!\!\! (-1)^j \#\, c_{2j}
\\
\hline
\end{array}
\end{equation*}
\vspace{-0.25cm}
\caption{Infinite table of positive quantities in higher dimensions. $\#$ stands for positive rational numbers and ✗ indicates  that positive quantities at that order in derivative cannot be constructed anymore.}
\label{tab:positivity}
\end{table}

\section{Large-$d$ behavior and saddle-point approximation}
\label{app:larged}

A direct way to motivate the normalization \eqref{eq:c0-c2-def} in the main text comes from considering the partial wave expansion \eqref{eq:PWE}. In \eqref{eq:PWE}, the main source of scaling with $d$ comes from $n_J^{(d)}$: both the partial waves and the Gegenbauer polynomials are bounded quantities for physical kinematics, but also at the crossing symmetric point. Ignoring the milder $J$-dependence, we have that $n_J^{(d)}\sim \Gamma(d/2)$, hence at large $d$, the amplitude grows factorially with $d$. Reciprocally, the partial-wave projection~\eqref{eq:PWProj} is consistent with that scaling, since $\mathcal{N}_d\sim 1/\Gamma(d/2)$ within the same approximation.

A more rigorous way to see this scaling uses partial wave unitarity, and a large-$d$ saddle of the partial wave projection coming from the integration measure $(1-z^2)^{(d-4)/2}$. It localities on a saddle at $z=0$ ($\pi/2$ orthogonal scattering), for which $f_0(s) =\frac{ \mathcal{N}_d}{2\sqrt{2\pi d}} A(s,2-s/2)$. 
Partial wave unitarity then rigorously bounds the value of the amplitude at $\theta=\pi/2$ for physical $s\geq 4$: 
$|1+i \phi^2_d(s) \frac{ \mathcal{N}_d}{2\sqrt{2\pi d}} A(s,2-\frac s2)|\leq1$, 
where as usual it is understood that the amplitude is evaluated slightly above the real axis at $s = \Re(s)+i 0^+$. The bound is compatible with the simple reasoning above and we expect that:
\begin{equation}
\forall d, \quad \mathcal{N}_d\times A(s,t) \sim O(1)\,,
\end{equation}
in the physical regime, where partial-wave unitarity holds, i.e. $s\geq 4, t<0$.
For non-zero $J$, the same reasoning can be done. The Gegenbauer now contribute to the saddle which becomes $J$-dependent, since $P_J^{(d\rightarrow\infty)}(z) \sim z^J$, and we find a similar localization to $z=\frac{\sqrt{J}}{\sqrt{d+J-4}}$
\begin{equation}
f_J(s) \underset{d\to\infty}{=} \frac{ \mathcal{N}_d}{2\sqrt{2\pi d}}d^{-J/2} (e^{-J/2}) J^{J/2} A\left(s,\tfrac12(4-s)\left(1 -\sqrt{\tfrac J D} \right)\right)\,.
\end{equation}
Thanks of the factor $d^{-J/2}$, if we could exchange the infinite-$d$ limit and the partial wave expansion, we would conclude that the amplitude is dominated solely by the spin-$0$ term. But the operations do not commute (as is obvious looking at the  divergent term $J^{J/2}$ which prevents to apply standard dominated convergence theorems), and this makes the infinite $d$ limit not trivial. Physically, fixed $J$ large $d$ probes only $\theta=\pi/2$ region, and the scattering (and crossing symmetry) is dominated by parametrically large spins, for which $J\sim d$ and also the tail $J\gg d$ contribute. This regime is much more difficult to control because of the Gegenbauer's asymptotics. In addition, since ACU  might just not make sense at arbitrary large $d$, especially $d>11$, it is unclear if such a construction would be useful anyway. We will investigate this further somewhere else.

There is a similar story in the $d=2$ limit, where the amplitude reduces to its forward limit $z=\pm1$, i.e., $\theta=\pm\pi$, reading $f_0(s) \sim A(s,4-s)$, explained in Appendix B of \cite{Chen:2022nym} to see how amplitude localizes at $\cos \theta = \pm 1$. The difference being that in $d=2$, all partial waves except spin-zero turn off and the problem becomes exactly solvable~\cite{Paulos:2016but}. Moreover, the expansion around 
$s=t=4/3$ becomes ill-defined, so do the definition of the Wilson coefficients we use in this letter.